\documentclass[aps,prl,twocolumn,superscriptaddress,longbibliography,nofootinbib]{revtex4-1}

\usepackage[utf8]{inputenc}
\usepackage{graphicx}
\usepackage{hyperref}
\usepackage{xcolor}

\usepackage{amsmath,amsthm,amssymb}
\usepackage{braket}
\usepackage{bbm,bm}
\usepackage[bb=boondox]{mathalfa}

\newcommand{\ie}{{\it i.e.},\ }

\newcommand{\id}{\mathbb{1}}
\newcommand{\Tr}{\operatorname{Tr}}
\newcommand{\ii}{\operatorname{i}}
\renewcommand{\Im}{\operatorname{Im}}
\renewcommand{\Re}{\operatorname{Re}}

\begin{document}
\title{The Page Curve for Fermionic Gaussian States}

\author{Eugenio Bianchi}
\email{ebianchi@psu.edu}
\affiliation{Department of Physics, The Pennsylvania State University, University Park, PA 16802, USA}
\affiliation{Institute for Gravitation and the Cosmos, The Pennsylvania State University, University Park, PA 16802, USA}

\author{Lucas Hackl}
\email{lucas.hackl@unimelb.edu.au}
\affiliation{School of Mathematics and Statistics, The University of Melbourne, Parkville, VIC 3010, Australia}
\affiliation{School of Physics, The University of Melbourne, Parkville, VIC 3010, Australia}

\author{Mario Kieburg}
\email{m.kieburg@unimelb.edu.au}
\affiliation{School of Mathematics and Statistics, The University of Melbourne, Parkville, VIC 3010, Australia}

\begin{abstract}
In a seminal paper, Page found the exact formula for the average entanglement entropy for a pure random state. We consider the analogous problem for the ensemble of pure fermionic Gaussian states, which plays a crucial role in the context of random free Hamiltonians. Using recent results from random matrix theory, we show that the average entanglement entropy of pure random fermionic Gaussian states in a subsystem of $N_A$ out of $N$ degrees of freedom is given by $\langle S_A\rangle_\mathrm{G}\!=\!(N\!-\!\tfrac{1}{2})\Psi(2N)\!+\!(\tfrac{1}{4}\!-\!N_A)\Psi(N)\!+\!(\tfrac{1}{2}\!+\!N_A\!-\!N)\Psi(2N\!-\!2N_A)\!-\!\tfrac{1}{4}\Psi(N\!-\!N_A)\!-\!N_A$, where $\Psi$ is the digamma function. Its asymptotic behavior in the thermodynamic limit is given by $\langle S_A\rangle_\mathrm{G}\!=\! N(\log 2-1)f+N(f-1)\log(1-f)+\tfrac{1}{2}f+\tfrac{1}{4}\log{(1-f)}\,+\,O(1/N)$, where $f=N_A/N$. Remarkably, its leading order agrees with the average over eigenstates of random quadratic Hamiltonians with number conservation, as found by {\L}yd{\.z}ba, Rigol and Vidmar. Finally, we compute the variance in the thermodynamic limit, given by the constant $\lim_{N\to\infty}(\Delta S_A)^2_{\mathrm{G}}=\frac{1}{2}(f+f^2+\log(1-f))$.
\end{abstract}

\maketitle

\emph{Introduction}.---Entanglement is a hallmark of quantum theory \cite{bell1964einstein,bell1966problem}. The study of the von Neumann bipartite entanglement entropy plays a central role in the quantum foundations of statistical mechanics
\cite{deutsch_91,srednicki_94,rigol_dunjko_08,d2016quantum, gogolin2016equilibration,deutsch2018eigenstate,goldstein_lebowitz_06, popescu_short_06,tasaki_98,polkovnikov2011colloquium,vidmar2017entanglement,liu2018quantum,vidmar2018volume,hackl2019average,Vidmar:2017pak,bianchi2019typical,lydzba2020eigenstate,lydzba2021entanglement,bernard2021entanglement}, in quantum information theory \cite{Hayden:2006,Hayden:2007cs,Sekino:2008he,Hosur:2015ylk,Roberts:2016hpo,Fujita:2017pju,Lu:2017tbo,Fujita:2018wtr}, in the formulation of the black hole information puzzle \cite{Page:1993wv,Giddings:2012bm,Braunstein:2009my,Almheiri:2012rt,Marolf:2017jkr,Harlow:2014yka,Bianchi:2014bma,Abdolrahimi:2015tha}, and the study of the quantum nature of spacetime geometry \cite{VanRaamsdonk:2010pw,Bianchi:2012ev,Jacobson:2015hqa,Bianchi_2019,Baytas:2018wjd,Qi:2018ogs}. Also experimentally there has been recently tremendous progress in measuring entanglement entropy in optical lattices with ultracold atoms~\cite{greiner_mandel_02b}. 

\begin{figure}
    \centering
    \includegraphics[width=0.48\textwidth]{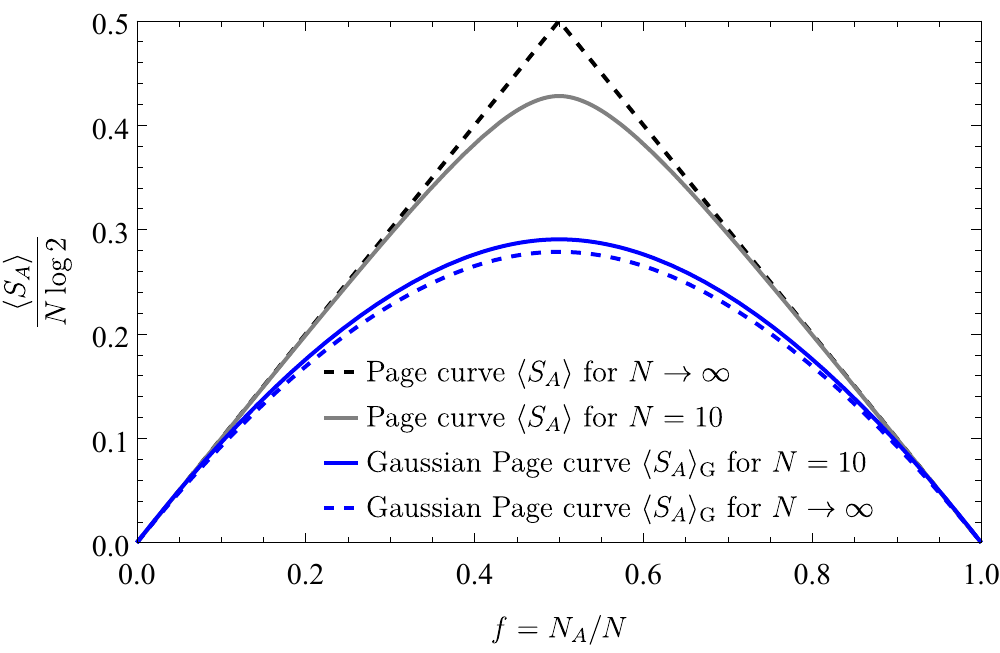}
    \caption{We compare the Page curve for random states $\braket{S_A}$ to the Page curve for random \emph{Gaussian} states $\braket{S_A}_\mathrm{G}$ for a system with $N$ fermionic degrees of freedom.}
    \label{fig:page-curve}
\end{figure}

In a seminal paper \cite{page1993average}, Page showed that, when an isolated quantum system is in a random pure state, the average entanglement entropy of a subsystem is close to maximal. In particular, he conjectured an exact formula for the average, taken with respect to the Haar measure over states in a finite-dimensional Hilbert space. In this letter we address the analogous problem for the ensemble of pure fermionic Gaussian states. We compute the average entanglement entropy $\braket{S_A}_\mathrm{G}$ of those states \eqref{eq:main-result} and study its properties (figure~\ref{fig:page-curve}) with the help of random matrix theory.

Pure fermionic Gaussian states appear as ground states and  eigenstates of free, \ie quadratic, Hamiltonians, and remain Gaussian in the time evolution after a free quantum quench~\cite{fagotti2008evolution,alba2018entanglement}. They play an important role in quantum computing in the context of matchgates~\cite{valiant2002quantum}. Moreover, there has been an increased interest in fermionic Gaussian states from the perspective of quantum chaos~\cite{srednicki_94}, and the eigenstate thermalization hypothesis~\cite{rigol_dunjko_08,vidmar16,magan2016random}.  The average eigenstate entanglement is of particular interest in this context~\cite{vidmar2017entanglement,vidmar2018volume,hackl2019average}, where one averages the entanglement entropy over the discrete set of eigenstates, which are Gaussian states for a given quadratic Hamiltonian. Our main results, the average entropy~\eqref{eq:main-result} and particularly its thermodynamic limit~\eqref{eq:thermodynamic-limit}, unveil close relations to recent work on the average entanglement entropy of eigenstates of quadratic Hamiltonians\ \cite{liu2018quantum,lydzba2020eigenstate,lydzba2021entanglement}. In particular, our formula~\eqref{eq:thermodynamic-limit} coincides in the thermodynamic limit with the average of the entanglement entropy with respect to the eigenstates of random quadratic Hamiltonians with number conservation (later numerically confirmed in\ \cite{lydzba2021entanglement} to also apply to random Hamiltonians without number conservation).

\medskip

\emph{The Page curve}.--- Before we come to the fermionic Gaussian states, let us briefly recall Page's result. In a quantum system consisting of $N$ spin $1/2$ fermions, a subsystem of $N_A$ fermions (with $N=N_A+N_B$ and $N_A\leq N_B$) defines a bipartition of the Hilbert space of states as $\mathcal{H}=\mathcal{H}_A\otimes\mathcal{H}_B$, with dimensions $\dim\mathcal{H}_A=2^{N_A}$ and $\dim\mathcal{H}_B=2^{N_B}$. Given a pure state $\ket{\psi}\in \mathcal{H}$, the entanglement entropy of the subsystem is $S_A(\ket{\psi})=-\Tr(\rho_A\log\rho_A)$ with $\rho_A=\Tr_B\ket{\psi}\!\bra{\psi}$ the induced density operator where the other $N_B$ fermions are traced out. The average over all states in $\mathcal{H}$ is
\begin{equation}
\braket{S_A}=\int\! d\mu(\psi)\, S_A(\ket{\psi})=\int_{U\in U(2^N)} \hspace{-3em}dU\;S_A(U\!\ket{0})\,,
\label{eq:page-average}
\end{equation}
where $d\mu(\psi)$ is the uniform measure in $\mathcal{H}$. This uniform measure on the $2^{2N}-1$ dimensional sphere can be obtained by fixing an arbitrary reference state $\ket{0}$ and acting on it with a unitary transformation $U$ distributed uniformly with respect to the Haar measure $dU$ over the unitary group $U\in U(2^N)$. In \cite{page1993average}, Page conjectured the formula (later proven in \cite{foong1994proof,sanchez1995simple,Sen:1996ph})
\begin{align}
\textstyle  \braket{S_A}=\Psi(2^N+1)-\Psi(2^{N-N_A}+1)-\frac{2^{N_A}-1}{2^{N-N_A+1}}\,,\label{eq:page-formula}
\end{align}
where $\Psi(z)=\Gamma'(z)/\Gamma(z)$ is the digamma function.

In the thermodynamic limit $N\to \infty$ with finite subsystem fraction $f=N_A/N\leq1/2$, the average entropy reduces to
\begin{equation}
\textstyle \braket{S_A}\sim f N \log 2\,-\,\frac{1}{2}\,\mathrm{e}^{\,-(1-2f)N\log 2}\,.
\label{eq:page-thermodynamic}
\end{equation}
Thence, for $f<1/2$, the average entanglement entropy approaches exponentially the entropy of a maximally mixed state. Similarly, in the thermodynamic limit, the dispersion around the average \cite{vivo_pato_16,wei2017proof,bianchi2019typical}  scales as
\begin{equation}
\textstyle\Delta S_A\sim\begin{cases}
\; 2^{-(1-f) N-\frac{1}{2}} & 0<f<\frac{1}{2}\\[.5em]
\; 2^{-\frac{1}{2}N-1} & f=\frac{1}{2}
\end{cases}\label{eq:variance-page}
\end{equation}
and vanishes exponentially. Consequently, a typical state in the Hilbert space is extremely close to being maximally entangled.

Let us turn our focus to fermionic Gaussian states which are states annihilated by a set of fermionic annihilation operators. Given a reference Gaussian state $\ket{J_0}$, all Gaussian states $\ket{J_M}$ can be generated via Bogoliubov transformations. These states form a submanifold in the manifold of pure states. The uniform measure $d\mu_\mathrm{G}(J)$ over Gaussian states can be defined in terms of the Haar measure over Bogoliubov transformation, \ie real orthogonal transformations $M\in O(2N)$~\cite{hackl2020bosonic}. The average entanglement entropy over fermionic Gaussian states is then
\begin{equation}
\braket{S_A}_\mathrm{G}=\int\!\! d\mu_\mathrm{G}(J)\, S_A(\ket{J})=\!\int_{M\in O(2N)}\hspace{-3em} dM\;S_A(\ket{J_M})\,.
\label{eq:gaussian-average}
\end{equation}
Using random matrix theory, we derive the following exact formula for the average entanglement entropy:
\begin{align}
\begin{split}
\textstyle \hspace{-2mm} \braket{S_A}_\mathrm{G}&=(N\!-\!\tfrac{1}{2})\Psi(2N)+(\tfrac{1}{2}\!+\!N_A\!-\!N)\Psi(2N\!-\!2N_A)\\[.5em]
    &\quad +(\tfrac{1}{4}\!-\!N_A)\Psi(N)-\tfrac{1}{4}\Psi(N\!-\!N_A)-N_A\,.
\end{split}\label{eq:main-result}
\end{align}
In the thermodynamic limit $N\to\infty$ with finite fraction $ f=N_A/N\leq 1/2$, a series expansion in $N$ yields
\begin{align}
\begin{split}
\braket{S_A}_\mathrm{G}&\sim N\,\big((\log{2}\!-\!1)f+\!(f\!-\!1)\log(1\!-\!f)\big)\\[.5em]
&\quad +\tfrac{1}{2}f+\tfrac{1}{4}\log{(1-f)}\,+\,O(1/N)\,,
\end{split}
\label{eq:thermodynamic-limit}
\end{align}
whose leading order term agrees with the expression deduced in\ \cite{lydzba2020eigenstate} for the average over a different set of states, as we will review in our discussion. We also find that the standard deviation approaches the constant
\begin{equation}
\lim_{N\to\infty}(\Delta S_A)_{\mathrm{G}}=\sqrt{\frac{f+f^2+\log(1-f)}{2}}\,.
\end{equation}
We outline the derivation of these results in the ensuing discussion.

\medskip

\emph{Average entropy}.---A quantum system with $N$ fermionic degrees of freedom can be formulated in terms of a set of creation and annihilation operators $\hat{a}_i^\dagger$ and $\hat{a}_i$ with canonical anti-commutation relations, $\{\hat{a}_i,\hat{a}_j^\dagger\}=\delta_{ij}$, $\{\hat{a}_i,\hat{a}_j\}=0$ and $i,j=1\ldots N$. Equivalently, we can introduce $2N$ Majorana modes $\hat{\xi}_\mu$ with $\mu=1\ldots 2N$ and
\begin{align}
\textstyle    \hat{\xi}_i=\frac{1}{\sqrt{2}}(\hat{a}^\dagger_i+\hat{a}_i)\quad\text{and}\quad\hat{\xi}_{N+i}=\frac{\ii}{\sqrt{2}}(\hat{a}^\dagger_i-\hat{a}_i)\,.\label{eq:Majorana-modes}
\end{align}
A Bogoliubov transformation $\hat{a}'_i=\sum^N_{j=1}(\alpha_{ij}\hat{a}_j+\beta_{ij}\hat{a}_j^\dagger)$ transforms the operators $\hat{\xi}_\mu$ to the new ones $\hat{\xi}_\mu'$ according to $\hat{\xi}_\mu'=\sum^{2N}_{\nu=1}M_{\mu\nu}\,\hat{\xi}_\nu$, where the $2N\times 2N$ matrix $M_{\mu\nu}$ is given by \cite{windt2020local}
\begin{align}
    M=\begin{pmatrix}
    \Re(\alpha+\beta) & \Im(\beta-\alpha)\\
    \Im(\alpha+\beta) & \Re(\alpha-\beta)
    \end{pmatrix}\,.
\end{align}
The requirement that the anti-commutation relations are preserved is equivalent to the condition $MM^\intercal=\id$, \ie $M$ must be an orthogonal matrix in $O(2N)$.

To define the uniform average over fermionic Gaussian states, we exploit the notion of a complex structure $J$ (\ie $J^2=-1$) and its relation to the correlation function~\cite{hackl2020bosonic}. The starting point is that a fermionic Gaussian state is defined as the ground state of a set of annihilation operators. We call $\ket{J_0}$ the state annihilated by the reference operators $\hat{a}_i$ and $\ket{J}$ the state annihilated by the Bogoliubov-transformed operators  $\hat{a}'_i$, \ie $\hat{a}'_i |J\rangle=0$. The label $J$ stands for the matrix $J=MJ_0M^{-1}$ determined by the expectation value of the commutator of two Majorana modes~\cite{hackl2020bosonic},
\begin{align}
\braket{J|\,[\hat{\xi}_\mu,\hat{\xi}_\nu]\,|J}=  \ii\,  J_{\mu\nu}\quad \text{with}\quad J_0=\left(\begin{array}{c|c}
     0    & \id \\
         \hline
    -\id     & 0
    \end{array}\right)\,.
\end{align}
The entanglement entropy of a Gaussian state $\ket{J}$ is directly related to the spectrum of the $2N_A\times 2N_A$ left-upper sub-block $[J]_A$ of $J$ via the formula  $S_A(\ket{J})=\sum^{N_A}_{i=1}s(x_i)$ with \cite{Peschel2003,Peschel2009,hackl2020bosonic,windt2020local}
\begin{align}
    s(x)=-\left(\tfrac{1-x}{2}\right)\log\left(\tfrac{1-x}{2}\right)-\left(\tfrac{1+x}{2}\right)\log\left(\tfrac{1+x}{2}\right)\,,\label{eq:entropy-formula}
\end{align}
where $x_i\in[0,1]$ are the singular values of $[J]_A$.

Having defined fermionic Gaussian states in terms of a reference state $\ket{J_0}$ and an orthogonal matrix $M$, we can express the uniform measure over Gaussian states in terms of the Haar measure over $O(2N)$ and compute the average entanglement entropy of Gaussian states exploiting~\eqref{eq:gaussian-average}. What we need to derive first is the joint probability distribution of the singular values $x_i$ of $[J]_A$. For this purpose we make repetitive use of~\cite[Proposition A.2]{kieburg2019multiplicative} by projecting away always two rows of the matrix $J$; first to $[J]_{N-1}$, then $[J]_{N-2}$ until we arrive at $[J]_{N_A}$. This yields for $x=(x_1,\dots,x_{N_A})$ the distribution
\begin{align}
  \textstyle  P(x)=\frac{\left(\det X\right)^2}{N_A!}\Big(\prod^{N_A-1}_{j=0}c_j^{-1}(1-x_{j+1}^2)^{\Delta}\Big)\,,\label{eq:probability}
\end{align}
where we have the $N_A\times N_A$ matrix $X$ and $c_j$,
\begin{align}
    X_{ij}&=p_{j-1}(x_i)=\mathcal{P}^{(\Delta,\Delta)}_{2j-2}(x_i)\,,\\
    c_{j}&=\frac{2^{2\Delta}\,[(2j+\Delta)!]^2}{(2j)!\,(2j+2\Delta)!\,(4j+2\Delta+1)}\,,\\
    \Delta&=N_B-N_A\geq0\,.
\end{align}
The functions $\mathcal{P}^{(\alpha,\beta)}_n(z)$ are the Jacobi polynomials and the distribution~\eqref{eq:probability} is related to the Jacobi ensemble~\cite{Forrester_2010}, one of the classical random matrix ensembles that can arise in various ways.

The $k$-point correlation functions~\cite{Forrester_2010} encode the whole spectral statistics of a random matrix,
\begin{align}
\textstyle R_k(x_1,\dots, x_k)&=\int P(x_1,\dots,x_k,y_1,\dots, y_{N_A-k}) d^{N_A-k}y\nonumber\\
&=\frac{(N_A-k)!}{N_A!}\det K(x_a,x_b)\,,\label{eq:Rk}
\end{align}
where $K(x_a,x_b)$ refers to the $k\times k$ matrix (with $a,b=1,\dots,k$) given by~\cite{Forrester_2010}
\begin{align}
\begin{split}
\hspace{-3mm}K(x,y)=\!\!\sum^{N_A-1}_{j=0}\!\psi_j(x)\psi_j(y),\,\,\,
\psi_j(x)=\tfrac{(1-x^2)^{\Delta/2}}{\sqrt{c_j}} p_j(x)
\end{split}\label{eq:Kmatrix}
\end{align}
with $\int_0^1 \psi_j(x)\psi_k(x)dx=\delta_{jk}$. Then, the level density is $\rho(x)=R_1(x)=\frac{1}{N_A}K(x,x)=\frac{1}{N_A}\sum^{N_A-1}_{j=0}\psi^2_j(x)$.

The average entropy is then given by the integral
\begin{align}
    \braket{S_A}_{\mathrm{G}}=N_A\int_0^1 s(x) \rho(x)
\end{align}
which can be evaluated by computing
\begin{align}
    I_\epsilon=\int^1_{-1}\left(\tfrac{x(1-x)^\epsilon}{2}-\tfrac{(1-x^2)^\epsilon}{4}\right)\rho(|x|)\,dx\,,
    %I^{(2)}_\epsilon=\int^1_{0}\frac{(1-x^2)^\epsilon}{2}\rho(x)\,dx\,,
\end{align}
such that $\braket{S_A}_\mathrm{G}=N_A\,(\partial_{\epsilon}I_\epsilon+\log{2})_{\epsilon\to 0}$. We combine the Jacobi polynomials with the other terms in the integrand whose integrals altogether give ratios of Gamma functions which, after the derivative yield the digamma functions in~\eqref{eq:main-result}.

We can compare the Gaussian Page curve~\eqref{eq:main-result} with the original Page curve~\eqref{eq:page-formula}, as illustrated in figure~\ref{fig:page-curve}. In the Gaussian case the thermodynamic limit is approached from above, while the original Page curve is approached from below. In fact, we can compute $\braket{S_A}=1/3$ and $\braket{S_A}_\mathrm{G}=1/2$ for $N_A=1$ with $N=2$, which shows that for small $N$ the average entanglement entropy of Gaussian states is \emph{above} the one of all states. This is in stark contrast to the thermodynamic limit, where the average entanglement entropy of Gaussian states is almost half of the one for all states.

\begin{figure*}[htb]
    \centering
    \includegraphics[width=\textwidth]{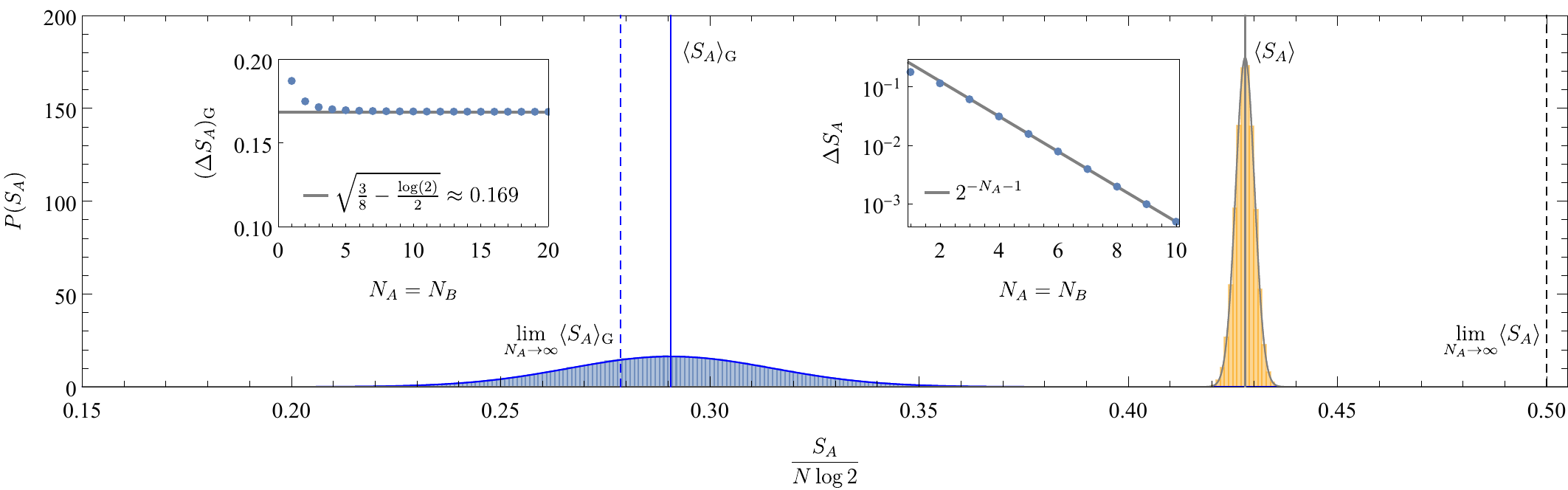}
    \caption{We compare the probability distributions $P(S_A)$ of the entanglement entropy $S_A$ for Gaussian states and general states in a fermionic system with $N_A=N_B=5$. The expectation value $\braket{S_A}$ and $\braket{S_A}_\mathrm{G}$ for $N_A=N_B$ and their thermodynamic limits are indicated by  vertical lines and vertical dashed lines, respectively. The insets depict the scaling of the standard deviation $\Delta S_A$ and $(\Delta S_A)_\mathrm{G}$ for $N_A=N_B\to\infty$.}
    \label{fig:prob}
\end{figure*}

\medskip

\begin{figure}[htb]
    \centering
    \includegraphics[width=.48\textwidth]{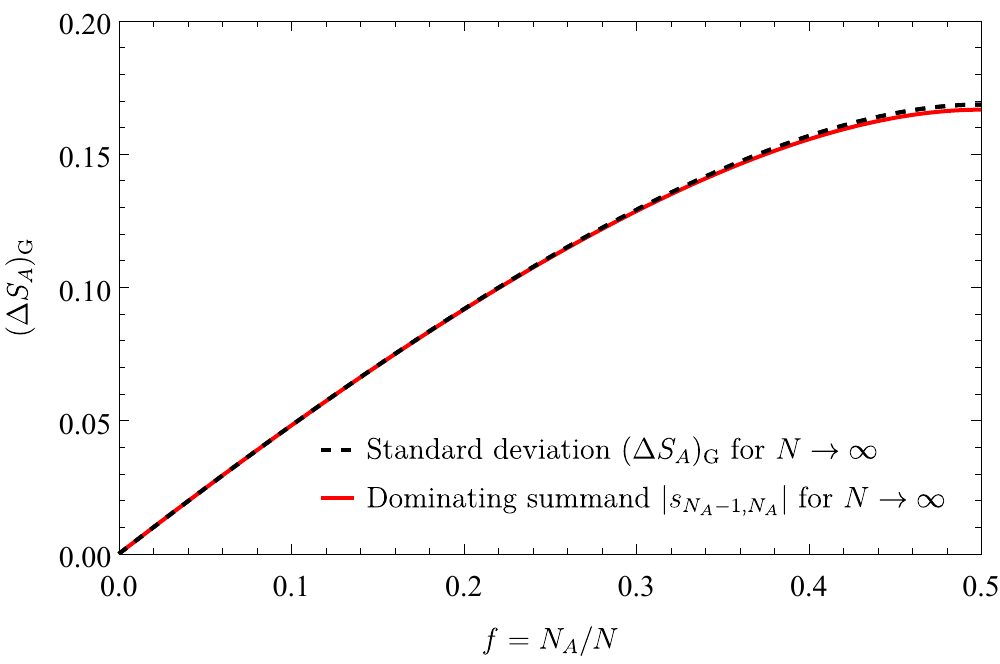}
    \caption{We compare the standard deviation $\lim_{N\to\infty}(\Delta S_A)_{\mathrm{G}}$  from~\eqref{eq:variance-sum} with the leading summand $|\overline{s}_{0,0}|\!=\!\lim_{N\to\infty}|s_{N_A-1,N_A}|\!=\!\frac{(3-4f)f}{6(1-f)}$ from~\eqref{eq:sbar-lk}.}
    \label{fig:peaked}
\end{figure}

\emph{Variance}.---An important question in the context of computing the average entanglement entropy is if this average is also typical, \ie if almost all states have an entanglement entropy close to the average as we take the thermodynamic limit. To answer this question, we compute the variance $(\Delta S_A)^2_\mathrm{G}=\braket{S_A^2}_\mathrm{G}-\braket{S_A}^2_\mathrm{G}$ of the probability distribution. For this, it is useful to define
\begin{align}
\begin{split}
    s_{ij}&=\int^1_0 s(x) \psi_i(x)\psi_j(x) dx\\
    &=-\left(\partial_{\epsilon}\int^{1}_{-1}\left(\frac{1-x}{2}\right)^{1+\epsilon}\psi_i(x)\psi_j(x)dx\right)_{\epsilon\to 0}\label{eq:sij-trick}\,,
\end{split}
\end{align}
where we used $\psi_i(x)=\psi_i(-x)$ to produce the two terms in~\eqref{eq:entropy-formula} by integrating over $[-1,1]$. We can interpret $s_{ij}$ as the matrix elements of the operator $s(x)$ with respect to the orthonormal basis  $\psi_j(x)$ in $[0,1]$. With this, we find
\begin{widetext}
\begin{align}
    (\Delta S_A)^2_{\mathrm{G}}&=\int_0^1 s^2(x) K(x,x)dx-\int_0^1s(x_1)s(x_2) K^2(x_1,x_2)d^2x=\int_0^1 s(x_1)s(x_2) K(x_1,x_2)(\delta(x_1-x_2)-K(x_1,x_2))d^2x\nonumber\\
    &=\int_0^1 s(x)s(y) \left(\sum^{N_A-1}_{i=0}\psi_i(x_1)\psi_i(x_2)\right)\left(\sum^{\infty}_{j=N_A}\psi_j(x_1)\psi_j(x_2)\right) d^2x=\sum^{N_A-1}_{i=0}\sum^\infty_{j=N_A}s^2_{ij}\quad\text{with}\label{eq:variance-sum}\\
    s^2_{ij}&=\tfrac{(2 j)! (2 \Delta +4 i+1) (\Delta +j+1) (2 \Delta +2 j+1) (2 \Delta +4 j+1) (2 (\Delta +i))! \left((1+\Delta-2 \Delta ^2) i-2 (\Delta -1) i^2+(\Delta +1) (2 j+1) (\Delta +j)\right)^2}{2 (2 i)! (2 i-2 j+1)^2 (i-j)^2 (-2 i+2 j+1)^2 (2 (\Delta +j+1))! (\Delta +i+j)^2 (\Delta +i+j+1)^2 (2 \Delta +2 i+2 j+1)^2}\,\,\,\text{for}\,\,\, i<j\,,\label{eq:sij-value}
\end{align}
\end{widetext}
where~\eqref{eq:sij-value} is only valid for $i<j$, which is all we need for the sum in~\eqref{eq:variance-sum}. Despite all terms in the sum of~\eqref{eq:variance-sum} are non-zero for large $N$, it is dominated by the summand $s^2_{N_A-1,N_A}$ (see figure~\ref{fig:peaked}, where we compare the sum vs. this dominating summand), so that it makes sense to consider the limit

\begin{align}
\begin{split}
    \overline{s}^2_{lk}&=\lim_{N\to\infty}s^2_{N_A-1-l,N_A+k}\\
    &=\tfrac{(\frac{1}{f}-1)^{-2 (k +l +1)} (2 k +2 l +3-4 f (k +l +1))^2}{4 (k +l +1)^2 (2 k +2 l +1)^2 (2 k +2l +3)^2}
    \end{split}\label{eq:sbar-lk}
\end{align}
with fixed $f=N_A/N$. From this, we find the variance
\begin{align}
    \hspace{-2mm}\lim_{N\to\infty}(\Delta S_A)^2_{\mathrm{G}}=\sum^\infty _{l,k=0}\overline{s}^2_{lk}=\frac{f+f^2+\log(1-f)}{2}\,,\label{eq:variance-sum}
\end{align}
where we could evaluate the sum analytically. That the variance approaches a constant is in line with numerical findings in~\cite{lydzba2020eigenstate,lydzba2021entanglement} and analytical studies of Renyi entropies~\cite{bernard2021entanglement}. Recall that the Page variance $\Delta S_A$ from~\eqref{eq:variance-page} converges to zero (with a behavior that differs for $f=\frac{1}{2}$), while the Gaussian standard deviation $(\Delta S_A)_{\mathrm{G}}$ approaches a constant and only its relative dispersion $(\Delta S_A)_{\mathrm{G}}/\braket{S_A}_{\mathrm{G}}$ will behave as $1/N$. In contrast, the standard deviation for (Gaussian) eigenstates of translationally invariant quadratic Hamiltonians was found in~\cite{vidmar2017entanglement} to scale as $\sqrt{N}$ (relative dispersion scaling as $1/\sqrt{N}$), which thus differs from both the Gaussian behavior found here and Page's result.

The stark contrast of the behavior of the standard deviation for all states vs. Gaussian states, \ie exponential  vs. constant, is closely connected to the dimension of the respective family of states (see figure~\ref{fig:prob}). While the real dimension of the manifold of pure fermionic states scales as $2^{N}$, the manifold of pure fermionic Gaussian states consists of two disconnected components of dimension $N(N-1)$ each. This behavior can be understood via Dyson's Brownian motion where the number of eigenvalues of the underlying random matrix (exponential in $N_A$ for pure fermionic states and quadratic in $N_A$) is crucial for the rate of convergence.

\medskip

\emph{Relation to random Hamiltonians}.---So far, we adopted the perspective of studying properties of a given ensemble of quantum states, namely the family of fermionic Gaussian states, without asking in what physical system one may necessarily encounter them. This is also the perspective of Page's original paper~\cite{page1993average} where he considers the family of all pure states, without reference to a specific Hamiltonian. Remarkably, for Hamiltonians with local interactions, while the ground state is far from Page-typical as it generally satisfies an area law~\cite{eisert2010colloquium}, the entanglement entropy of energy eigenstates can be obtained from typicality arguments \cite{deutsch_91,srednicki_94,rigol_dunjko_08,d2016quantum, gogolin2016equilibration,deutsch2018eigenstate,goldstein_lebowitz_06, popescu_short_06, tasaki_98,polkovnikov2011colloquium,vidmar2017entanglement,vidmar2018volume,hackl2019average,Vidmar:2017pak,bianchi2019typical}. Similarly, it is instructive to investigate for which random Hamiltonians the resulting ground states constitute the considered ensemble of fermionic Gaussian states discussed here.

For this aim, we consider the most general quadratic Hamiltonian for $N$ fermionic degrees of freedom,
\begin{align}
\hspace{-2mm}    \hat{H}=\!\!\!\sum_{i,j=1}^N\!(A_{ij} \hat{a}^\dagger_i\hat{a}_j\!+\!B_{ij} \hat{a}^\dagger_i\hat{a}_j^\dagger\!+{\rm h.c.})%\!B^*_{ij} \hat{a}_i\hat{a}_j)
    =\!\!\sum^{2N}_{\mu,\nu=1}\!\! \ii  h_{\mu\nu}\,\hat{\xi}_\mu\hat{\xi}_\nu,
    \label{eq:Hquad}
\end{align}
where the $2N$ Majorana modes were introduced in~\eqref{eq:Majorana-modes} and $h_{\mu\nu}$ is an anti-symmetric matrix with real entries (as also considered in\ \cite{lydzba2021entanglement}). Any such antisymmetric matrix can be block-diagonalized by means of an orthogonal transformation $M_{\mu\nu}$, such that
\begin{align}
    MhM^{-1}={\bigoplus}^{N}_{i=1}\begin{pmatrix}
    0 & \omega_i\\
    -\omega_i & 0
    \end{pmatrix}\,,
\end{align}
where $\omega_i\geq0$ leading to $\hat{H}=\sum^N_{i=1}\omega_i\,(\hat{b}_i^\dagger\hat{b}_i-\frac{1}{2})$ with transformed creation and annihilation operators.

If we randomly generate the matrix entries of $h_{\mu\nu}$ with respect to some $\mathrm{O}(2N)$ invariant probability distribution, for instance a Gaussian distribution, the orthogonal transformation $M$ that diagonalizes it will be Haar distributed. Therefore, the resulting ground state  of $\hat{H}$ is the state annihilated by $\hat{b}_i$ and is distributed according to the ensemble of fermionic Gaussian states considered so far. Moreover, the excited energy eigenstates of this random Hamiltonian are also Gaussian states and distributed according to the same ensemble. Remarkably, the result does not depend on which specific choice of $\mathrm{O}(2N)$ invariant distribution we use to generate $\hat{H}$: in fact, only the one-particle spectrum $\omega_i$ depends on this choice, and the properties of the energy eigenstates are independent of the associated eigenvalues (as long as no degeneracies are present). Therefore, the eigenstates of \eqref{eq:Hquad} are distributed as random Gaussian states $\ket{J}$ with Haar measure $d\mu_\mathrm{G}(J)$ as in \eqref{eq:gaussian-average}.

This result provides an analytical derivation of the numerical evidence found by {\L}yd{\.z}ba, Rigol and Vidmar in \cite{lydzba2021entanglement} that in the thermodynamic limit the average entropy of eigenstates of a random Hamiltonian \eqref{eq:Hquad} is given by \eqref{eq:thermodynamic-limit}. Moreover, the argument above extents the result to systems of finite size: the average eigenstate entanglement entropy of $\mathrm{O}(2N)$-invariant random quadratic Hamiltonians is given by the exact analytic formula \eqref{eq:main-result}.

On the other hand, imposing further constraints on the Hamiltonian \eqref{eq:Hquad}, such as requiring it to be particle number preserving or translationally invariant, will result in a submanifold of the manifold of fermionic Gaussian states \eqref{eq:gaussian-average}. Therefore we cannot expect to find the same statistical properties (average, variance) for the entanglement entropy at finite system size. Yet, in the large $N$ limit, the average over eigenstates of number preserving Hamiltonians studied in \ \cite{liu2018quantum,lydzba2020eigenstate} leads to an average entanglement entropy that agrees with our result \eqref{eq:thermodynamic-limit} in the thermodynamic limit.

\medskip

\emph{Discussion}---The main result of this letter is the analytical expression~\eqref{eq:main-result}, which is the analogue of Page's result for the ensemble of fermionic Gaussian states for systems of finite size, and its large $N$ behavior~\eqref{eq:thermodynamic-limit}. The derivation was made possible by recent advances in random matrix theory\ \cite{kieburg2019multiplicative}, which bear promise to be also relevant for other ensembles of states. Our results enable us to deduce a number of interesting properties of the Page curve of fermionic Gaussian states: (a) The curve admits a closed form expression in terms of digamma functions from which finite size corrections to the thermodynamic limit can be extracted. (b) In contrast to Page’s typicality, for fermionic Gaussian states the thermodynamic limit is approached from above and only algebraically fast, rather than exponentially. (c) The variance approaches a constant at large $N$ rather than decaying exponentially as in Page's case. (d) Finally, our result shows that whenever the subsystem fraction is finite in the thermodynamic limit, the average entanglement entropy is smaller than the maximal value $f\log 2$, but approaches it as the subsystem fraction $f$ goes to zero.

Our proof helps to clarify the relationship to the average entanglement entropy of ground states and eigenstates of random quadratic Hamiltonians, namely that these averages coincide in the thermodynamic limit provided that the Hamiltonian is sufficiently random. Let us emphasize that the function
\begin{align}
  \lim_{N\to\infty} \textstyle \frac{1}{N}\braket{S_A}_\mathrm{G}=(\log 2\!-\!1)f+(f\!-\!1)\log(1\!-\!f)
  \label{eq:LRV}
\end{align}
was found by {\L}yd{\.z}ba, Rigol and Vidmar in~\cite{lydzba2020eigenstate} as an average over energy eigenstates of random Hamiltonians with number conservation (and later shown numerically\ \cite{lydzba2021entanglement} to also apply to eigenstates of random quadratic Hamiltonians without number conservation). In the thermodynamic limit, the associated level density of the matrix $[J]_A$ for a similar model was also found previously in~\cite{liu2018quantum}, from which the value $\lim_{N=2N_A\to\infty} \frac{1}{N}\braket{S_A}=\log{2}-\tfrac{1}{2}$ was computed. Both papers construct their family of states from number preserving quadratic Hamiltonians, namely either as the ground state of the SYK2 Hamiltonian~\cite{liu2018quantum} or as one of its eigenstates~\cite{lydzba2020eigenstate}. In both cases, the set of states is determined by the subgroup $\mathrm{U}(N)$ of number-preserving Bogoliubov transformations, which is only a submanifold of the $\mathrm{O}(2N)$ manifold of Gaussian states considered here. For $\mathrm{O}(2N)$-invariant random quadratic Hamiltonians, we find that the average eigenstate entanglement entropy is given by the analytic formula \eqref{eq:main-result} for systems of finite size. Explaining from general arguments why in the thermodynamic limit the average \eqref{eq:LRV} arises more generally, identifying what is the universality class and computing the finite size corrections to the average and variance for different classes of random Hamiltonians would be an interesting avenue for future work.

The random matrix  techniques used here can also be applied to derive similar ``Page-like curves'' for Renyi entropies~\cite{bernard2021entanglement} and other information-theoretic quantities. While here we focused on energy eigenstates and time-independent Hamiltonians, our results provide also a prediction for the value of the equilibrium entanglement entropy under unitary evolution generated by a random time-dependent quadratic Hamiltonian~\cite{fagotti2008evolution,nahum2017quantum,bauer2017stochastic}.

Another Page-like curve was considered recently~\cite{vidmar2017entanglement,vidmar2018volume,hackl2019average} in the context of translationally invariant quadratic Hamiltonians, for which the average entanglement entropy over all eigenstates was computed. While this average involves a discrete set of states which differs depending on the chosen Hamiltonians, numerical evidence for several classes of translationally invariant quadratic models suggested the conjecture that the resulting curve is actually universal~\cite{hackl2020bosonic} in the thermodynamic limit. The most compelling explanation for such a behavior relies again on random matrix theory and assumes that any such discrete set will ultimately sample from the Haar measure on the manifold of translationally invariant Gaussian states. It would therefore be a meaningful avenue to adapt the methods developed in this letter to derive similar analytical expressions for the average entanglement entropy of translationally invariant Gaussian states.

\begin{acknowledgments}
\medskip

\emph{Acknowledgments}.---We would like to thank Pietro Don\`a, Peter Forrester, Marcos Rigol and Lev Vidmar for inspiring discussions and comments on the manuscript. Special thanks goes to Lorenzo Piroli who pointed out an error in our analysis for the variance, which we subsequently corrected by deriving its exact form in the thermodynamic limit. LH gratefully acknowledges support by the Alexander von Humboldt Foundation. EB acknowledges support by the NSF via the Grant PHY-1806428 and by the John Templeton Foundation via the ID 61466 grant, as part of the “Quantum Information Structure of Spacetime (QISS)” project (\hyperlink{http://www.qiss.fr}{qiss.fr}). 

.
\end{acknowledgments}

\bibliographystyle{JHEP}
\bibliography{references}

\providecommand{\href}[2]{#2}\begingroup\raggedright\begin{thebibliography}{10}

\bibitem{bell1964einstein}
J.S.~Bell, \emph{On the {E}instein {P}odolsky {R}osen paradox},
  \href{https://doi.org/10.1142/9789812386540_0002}{\emph{Physics Physique
  Fizika} {\bfseries 1} (1964) 195}.

\bibitem{bell1966problem}
J.S.~Bell, \emph{On the problem of hidden variables in quantum mechanics},
  \href{https://doi.org/10.1142/9789812386540_0001}{\emph{Reviews of Modern
  Physics} {\bfseries 38} (1966) 447}.

\bibitem{deutsch_91}
J.M.~Deutsch, \emph{Quantum statistical mechanics in a closed system},
  \href{https://doi.org/10.1103/PhysRevA.43.2046}{\emph{Physical Review A}
  {\bfseries 43} (1991) 2046}.

\bibitem{srednicki_94}
M.~Srednicki, \emph{Chaos and quantum thermalization},
  \href{https://doi.org/10.1103/PhysRevE.50.888}{\emph{Physical Review E}
  {\bfseries 50} (1994) 888}
  [\href{https://arxiv.org/abs/cond-mat/9403051}{{\ttfamily
  cond-mat/9403051}}].

\bibitem{rigol_dunjko_08}
M.~Rigol, V.~Dunjko and M.~Olshanii, \emph{Thermalization and its mechanism for
  generic isolated quantum systems},
  \href{https://doi.org/10.1038/nature06838}{\emph{Nature} {\bfseries 452}
  (2008) 854} [\href{https://arxiv.org/abs/0708.1324}{{\ttfamily 0708.1324}}].

\bibitem{d2016quantum}
L.~D'Alessio, Y.~Kafri, A.~Polkovnikov and M.~Rigol, \emph{From quantum chaos
  and eigenstate thermalization to statistical mechanics and thermodynamics},
  \href{https://doi.org/10.1080/00018732.2016.1198134}{\emph{Advances in
  Physics} {\bfseries 65} (2016) 239}
  [\href{https://arxiv.org/abs/1509.06411}{{\ttfamily 1509.06411}}].

\bibitem{gogolin2016equilibration}
C.~Gogolin and J.~Eisert, \emph{Equilibration, thermalisation, and the
  emergence of statistical mechanics in closed quantum systems},
  \href{https://doi.org/10.1088/0034-4885/79/5/056001}{\emph{Reports on
  Progress in Physics} {\bfseries 79} (2016) 056001}
  [\href{https://arxiv.org/abs/1503.07538}{{\ttfamily 1503.07538}}].

\bibitem{deutsch2018eigenstate}
J.M.~Deutsch, \emph{Eigenstate thermalization hypothesis},
  \href{https://doi.org/10.1088/1361-6633/aac9f1}{\emph{Reports on Progress in
  Physics} {\bfseries 81} (2018) 082001}
  [\href{https://arxiv.org/abs/1805.01616}{{\ttfamily 1805.01616}}].

\bibitem{goldstein_lebowitz_06}
S.~Goldstein, J.L.~Lebowitz, R.~Tumulka and N.~Zangh\`\i{}, \emph{Canonical
  typicality},
  \href{https://doi.org/10.1103/PhysRevLett.96.050403}{\emph{Physical Review
  Letters} {\bfseries 96} (2006) 050403}
  [\href{https://arxiv.org/abs/cond-mat/0511091}{{\ttfamily
  cond-mat/0511091}}].

\bibitem{popescu_short_06}
S.~Popescu, A.J.~Short and A.~Winter, \emph{Entanglement and the foundations of
  statistical mechanics},
  \href{https://doi.org/dx.doi.org/10.1038/nphys444}{\emph{Nature Physics}
  {\bfseries 2} (2006) 754}
  [\href{https://arxiv.org/abs/quant-ph/0511225}{{\ttfamily
  quant-ph/0511225}}].

\bibitem{tasaki_98}
H.~Tasaki, \emph{From quantum dynamics to the canonical distribution: General
  picture and a rigorous example},
  \href{https://doi.org/10.1103/PhysRevLett.80.1373}{\emph{Physical Review
  Letters} {\bfseries 80} (1998) 1373}
  [\href{https://arxiv.org/abs/1709.06259}{{\ttfamily 1709.06259}}].

\bibitem{polkovnikov2011colloquium}
A.~Polkovnikov, K.~Sengupta, A.~Silva and M.~Vengalattore, \emph{Colloquium:
  Nonequilibrium dynamics of closed interacting quantum systems},
  \href{https://doi.org/10.1103/revmodphys.83.863}{\emph{Reviews of Modern
  Physics} {\bfseries 83} (2011) 863}
  [\href{https://arxiv.org/abs/1007.5331}{{\ttfamily 1007.5331}}].

\bibitem{vidmar2017entanglement}
L.~Vidmar, L.~Hackl, E.~Bianchi and M.~Rigol, \emph{Entanglement entropy of
  eigenstates of quadratic fermionic hamiltonians},
  \href{https://doi.org/10.1103/physrevlett.119.020601}{\emph{Physical Review
  Letters} {\bfseries 119} (2017) 020601}
  [\href{https://arxiv.org/abs/1703.02979}{{\ttfamily 1703.02979}}].

\bibitem{liu2018quantum}
C.~Liu, X.~Chen and L.~Balents, \emph{Quantum entanglement of the
  {S}achdev-{Y}e-{K}itaev models}, {\emph{Physical Review B} {\bfseries 97}
  (2018) 245126} [\href{https://arxiv.org/abs/1709.06259}{{\ttfamily
  1709.06259}}].

\bibitem{vidmar2018volume}
L.~Vidmar, L.~Hackl, E.~Bianchi and M.~Rigol, \emph{Volume law and quantum
  criticality in the entanglement entropy of excited eigenstates of the quantum
  ising model},
  \href{https://doi.org/10.1103/physrevlett.121.220602}{\emph{Physical Review
  Letters} {\bfseries 121} (2018) 220602}
  [\href{https://arxiv.org/abs/1808.08963}{{\ttfamily 1808.08963}}].

\bibitem{hackl2019average}
L.~Hackl, L.~Vidmar, M.~Rigol and E.~Bianchi, \emph{Average eigenstate
  entanglement entropy of the xy chain in a transverse field and its
  universality for translationally invariant quadratic fermionic models},
  \href{https://doi.org/10.1103/PhysRevB.99.075123}{\emph{Physical Review B}
  {\bfseries 99} (2019) 075123}
  [\href{https://arxiv.org/abs/1812.08757}{{\ttfamily 1812.08757}}].

\bibitem{Vidmar:2017pak}
L.~Vidmar and M.~Rigol, \emph{Entanglement entropy of eigenstates of quantum
  chaotic hamiltonians},
  \href{https://doi.org/10.1103/PhysRevLett.119.220603}{\emph{Physical Review
  Letters} {\bfseries 119} (2017) 220603}
  [\href{https://arxiv.org/abs/1708.08453}{{\ttfamily 1708.08453}}].

\bibitem{bianchi2019typical}
E.~Bianchi and P.~Don{\`a}, \emph{Typical entanglement entropy in the presence
  of a center: Page curve and its variance},
  \href{https://doi.org/10.1103/physrevd.100.105010}{\emph{Physical Review D}
  {\bfseries 100} (2019) 105010}
  [\href{https://arxiv.org/abs/1904.08370}{{\ttfamily 1904.08370}}].

\bibitem{lydzba2020eigenstate}
P.~{\L}yd{\.z}ba, M.~Rigol and L.~Vidmar, \emph{Eigenstate entanglement entropy
  in random quadratic hamiltonians},
  \href{https://doi.org/10.1103/physrevlett.125.180604}{\emph{Physical Review
  Letters} {\bfseries 125} (2020) 180604}
  [\href{https://arxiv.org/abs/2006.11302}{{\ttfamily 2006.11302}}].

\bibitem{lydzba2021entanglement}
P.~{\L}yd{\.z}ba, M.~Rigol and L.~Vidmar, \emph{Entanglement in many-body
  eigenstates of quantum-chaotic quadratic hamiltonians},
  \href{https://doi.org/10.1103/PhysRevB.103.104206}{\emph{Physical Review B}
  {\bfseries 103} (2021) 104206}
  [\href{https://arxiv.org/abs/2101.05309}{{\ttfamily 2101.05309}}].

\bibitem{bernard2021entanglement}
D.~Bernard and L.~Piroli, \emph{Entanglement distribution in the quantum
  symmetric simple exclusion process}, {\emph{arXiv preprint arXiv:2102.04745}
  (2021) } [\href{https://arxiv.org/abs/2102.04745}{{\ttfamily 2102.04745}}].

\bibitem{Hayden:2006}
P.~Hayden, D.W.~Leung and A.~Winter, \emph{Aspects of generic entanglement},
  \href{https://doi.org/10.1007/s00220-006-1535-6}{\emph{Communications in
  Mathematical Physics} {\bfseries 265} (2006) 95}
  [\href{https://arxiv.org/abs/quant-ph/0407049}{{\ttfamily
  quant-ph/0407049}}].

\bibitem{Hayden:2007cs}
P.~Hayden and J.~Preskill, \emph{{Black holes as mirrors: Quantum information
  in random subsystems}},
  \href{https://doi.org/10.1088/1126-6708/2007/09/120}{\emph{JHEP} {\bfseries
  09} (2007) 120} [\href{https://arxiv.org/abs/0708.4025}{{\ttfamily
  0708.4025}}].

\bibitem{Sekino:2008he}
Y.~Sekino and L.~Susskind, \emph{{Fast Scramblers}},
  \href{https://doi.org/10.1088/1126-6708/2008/10/065}{\emph{JHEP} {\bfseries
  10} (2008) 065} [\href{https://arxiv.org/abs/0808.2096}{{\ttfamily
  0808.2096}}].

\bibitem{Hosur:2015ylk}
P.~Hosur, X.-L.~Qi, D.A.~Roberts and B.~Yoshida, \emph{{Chaos in quantum
  channels}}, \href{https://doi.org/10.1007/JHEP02(2016)004}{\emph{JHEP}
  {\bfseries 02} (2016) 004}
  [\href{https://arxiv.org/abs/1511.04021}{{\ttfamily 1511.04021}}].

\bibitem{Roberts:2016hpo}
D.A.~Roberts and B.~Yoshida, \emph{{Chaos and complexity by design}},
  \href{https://doi.org/10.1007/JHEP04(2017)121}{\emph{JHEP} {\bfseries 04}
  (2017) 121} [\href{https://arxiv.org/abs/1610.04903}{{\ttfamily
  1610.04903}}].

\bibitem{Fujita:2017pju}
Y.O.~Nakagawa, M.~Watanabe, S.~Sugiura and H.~Fujita, \emph{{Universality in
  volume-law entanglement of scrambled pure quantum states}},
  \href{https://doi.org/10.1038/s41467-018-03883-9}{\emph{Nature
  Communications} {\bfseries 9} (2018) 1635}
  [\href{https://arxiv.org/abs/1703.02993}{{\ttfamily 1703.02993}}].

\bibitem{Lu:2017tbo}
T.-C.~Lu and T.~Grover, \emph{{Renyi Entropy of Chaotic Eigenstates}},
  \href{https://doi.org/10.1103/PhysRevE.99.032111}{\emph{Physical Review E}
  {\bfseries 99} (2019) 032111}
  [\href{https://arxiv.org/abs/1709.08784}{{\ttfamily 1709.08784}}].

\bibitem{Fujita:2018wtr}
H.~Fujita, Y.O.~Nakagawa, S.~Sugiura and M.~Watanabe, \emph{{Page Curves for
  General Interacting Systems}},
  \href{https://doi.org/10.1007/JHEP12(2018)112}{\emph{JHEP} {\bfseries 12}
  (2018) 112} [\href{https://arxiv.org/abs/1805.11610}{{\ttfamily
  1805.11610}}].

\bibitem{Page:1993wv}
D.N.~Page, \emph{Information in black hole radiation},
  \href{https://doi.org/10.1103/PhysRevLett.71.3743}{\emph{Physical Review
  Letters} {\bfseries 71} (1993) 3743}
  [\href{https://arxiv.org/abs/hep-th/9306083}{{\ttfamily hep-th/9306083}}].

\bibitem{Giddings:2012bm}
S.B.~Giddings, \emph{{Black holes, quantum information, and unitary
  evolution}}, \href{https://doi.org/10.1103/PhysRevD.85.124063}{\emph{Physical
  Review D} {\bfseries 85} (2012) 124063}
  [\href{https://arxiv.org/abs/1201.1037}{{\ttfamily 1201.1037}}].

\bibitem{Braunstein:2009my}
S.L.~Braunstein, S.~Pirandola and K.~\.Zyczkowski, \emph{{Better Late than
  Never: Information Retrieval from Black Holes}},
  \href{https://doi.org/10.1103/PhysRevLett.110.101301}{\emph{Physical Review
  Letters} {\bfseries 110} (2013) 101301}
  [\href{https://arxiv.org/abs/0907.1190}{{\ttfamily 0907.1190}}].

\bibitem{Almheiri:2012rt}
A.~Almheiri, D.~Marolf, J.~Polchinski and J.~Sully, \emph{{Black Holes:
  Complementarity or Firewalls?}},
  \href{https://doi.org/10.1007/JHEP02(2013)062}{\emph{JHEP} {\bfseries 02}
  (2013) 062} [\href{https://arxiv.org/abs/1207.3123}{{\ttfamily 1207.3123}}].

\bibitem{Marolf:2017jkr}
D.~Marolf, \emph{{The Black Hole information problem: past, present, and
  future}}, \href{https://doi.org/10.1088/1361-6633/aa77cc}{\emph{Reports on
  Progress in Physics} {\bfseries 80} (2017) 092001}
  [\href{https://arxiv.org/abs/1703.02143}{{\ttfamily 1703.02143}}].

\bibitem{Harlow:2014yka}
D.~Harlow, \emph{{Jerusalem Lectures on Black Holes and Quantum Information}},
  \href{https://doi.org/10.1103/RevModPhys.88.015002}{\emph{Reviews of Modern
  Physics} {\bfseries 88} (2016) 015002}
  [\href{https://arxiv.org/abs/1409.1231}{{\ttfamily 1409.1231}}].

\bibitem{Bianchi:2014bma}
E.~Bianchi, T.~De~Lorenzo and M.~Smerlak, \emph{{Entanglement entropy
  production in gravitational collapse: covariant regularization and solvable
  models}}, \href{https://doi.org/10.1007/JHEP06(2015)180}{\emph{JHEP}
  {\bfseries 06} (2015) 180} [\href{https://arxiv.org/abs/1409.0144}{{\ttfamily
  1409.0144}}].

\bibitem{Abdolrahimi:2015tha}
S.~Abdolrahimi and D.N.~Page, \emph{{Hawking Radiation Energy and Entropy from
  a Bianchi-Smerlak Semiclassical Black Hole}},
  \href{https://doi.org/10.1103/PhysRevD.92.083005}{\emph{Physical Review D}
  {\bfseries 92} (2015) 083005}
  [\href{https://arxiv.org/abs/1506.01018}{{\ttfamily 1506.01018}}].

\bibitem{VanRaamsdonk:2010pw}
M.~Van~Raamsdonk, \emph{Building up spacetime with quantum entanglement},
  \href{https://doi.org/10.1007/s10714-010-1034-0,
  10.1142/S0218271810018529}{\emph{General Relativity and Gravitation}
  {\bfseries 42} (2010) 2323}
  [\href{https://arxiv.org/abs/1005.3035}{{\ttfamily 1005.3035}}].

\bibitem{Bianchi:2012ev}
E.~Bianchi and R.C.~Myers, \emph{{On the Architecture of Spacetime Geometry}},
  \href{https://doi.org/10.1088/0264-9381/31/21/214002}{\emph{Classical and
  Quantum Gravity} {\bfseries 31} (2014) 214002}
  [\href{https://arxiv.org/abs/1212.5183}{{\ttfamily 1212.5183}}].

\bibitem{Jacobson:2015hqa}
T.~Jacobson, \emph{Entanglement equilibrium and the einstein equation},
  \href{https://doi.org/10.1103/PhysRevLett.116.201101}{\emph{Physical Review
  Letters} {\bfseries 116} (2016) 201101}
  [\href{https://arxiv.org/abs/1505.04753}{{\ttfamily 1505.04753}}].

\bibitem{Bianchi_2019}
E.~Bianchi, P.~Don{\`{a}} and I.~Vilensky, \emph{Entanglement entropy of
  bell-network states in loop quantum gravity: Analytical and numerical
  results}, \href{https://doi.org/10.1103/physrevd.99.086013}{\emph{Physical
  Review D} {\bfseries 99} (2019) 086013}
  [\href{https://arxiv.org/abs/1812.10996}{{\ttfamily 1812.10996}}].

\bibitem{Baytas:2018wjd}
B.~Baytas, E.~Bianchi and N.~Yokomizo, \emph{{Gluing polyhedra with
  entanglement in loop quantum gravity}},
  \href{https://doi.org/10.1103/PhysRevD.98.026001}{\emph{Phys. Rev.}
  {\bfseries D98} (2018) 026001}
  [\href{https://arxiv.org/abs/1805.05856}{{\ttfamily 1805.05856}}].

\bibitem{Qi:2018ogs}
X.-L.~Qi, \emph{{Does gravity come from quantum information?}},
  \href{https://doi.org/10.1038/s41567-018-0297-3}{\emph{Nature Physics}
  {\bfseries 14} (2018) 984}.

\bibitem{greiner_mandel_02b}
M.~Greiner, O.~Mandel, T.W.~H\"ansch and I.~Bloch, \emph{Collapse and revival
  of the matter wave field of a $\mathrm{B}$ose-$\mathrm{E}$instein
  condensate}, \href{https://doi.org/10.1038/nature00968}{\emph{Nature}
  {\bfseries 419} (2002) 51}.

\bibitem{page1993average}
D.N.~Page, \emph{Average entropy of a subsystem},
  \href{https://doi.org/10.1103/physrevlett.71.1291}{\emph{Physical Review
  Letters} {\bfseries 71} (1993) 1291}
  [\href{https://arxiv.org/abs/gr-qc/9305007}{{\ttfamily gr-qc/9305007}}].

\bibitem{fagotti2008evolution}
M.~Fagotti and P.~Calabrese, \emph{Evolution of entanglement entropy following
  a quantum quench: Analytic results for the x y chain in a transverse magnetic
  field}, \href{https://doi.org/10.1103/physreva.78.010306}{\emph{Physical
  Review A} {\bfseries 78} (2008) 010306(R)}
  [\href{https://arxiv.org/abs/0804.3559}{{\ttfamily 0804.3559}}].

\bibitem{alba2018entanglement}
V.~Alba and P.~Calabrese, \emph{Entanglement dynamics after quantum quenches in
  generic integrable systems},
  \href{https://doi.org/10.21468/scipostphys.4.3.017}{\emph{SciPost Physics}
  {\bfseries 4} (2018) 017} [\href{https://arxiv.org/abs/1712.07529}{{\ttfamily
  1712.07529}}].

\bibitem{valiant2002quantum}
L.G.~Valiant, \emph{Quantum circuits that can be simulated classically in
  polynomial time}, \href{https://doi.org/10.1137/s0097539700377025}{\emph{SIAM
  J Comput} {\bfseries 31} (2002) 1229}.

\bibitem{vidmar16}
L.~Vidmar and M.~Rigol, \emph{Generalized gibbs ensemble in integrable lattice
  models},
  \href{https://doi.org/10.1088/1742-5468/2016/06/064007}{\emph{Journal of
  Statistical Mechanics} {\bfseries 2} (2016) 064007}
  [\href{https://arxiv.org/abs/1604.03990}{{\ttfamily 1604.03990}}].

\bibitem{magan2016random}
J.M.~Mag{\'a}n, \emph{Random free fermions: An analytical example of eigenstate
  thermalization},
  \href{https://doi.org/10.1103/PhysRevLett.116.030401}{\emph{Physical Review
  Letters} {\bfseries 116} (2016) 030401}
  [\href{https://arxiv.org/abs/1508.05339}{{\ttfamily 1508.05339}}].

\bibitem{foong1994proof}
S.K.~Foong and S.~Kanno, \emph{Proof of page’s conjecture on the average
  entropy of a subsystem},
  \href{https://doi.org/10.1103/physrevlett.72.1148}{\emph{Physical Review
  Letters} {\bfseries 72} (1994) 1148}.

\bibitem{sanchez1995simple}
J.~S{\'a}nchez-Ruiz, \emph{Simple proof of page’s conjecture on the average
  entropy of a subsystem},
  \href{https://doi.org/10.1103/physreve.52.5653}{\emph{Physical Review E}
  {\bfseries 52} (1995) 5653}.

\bibitem{Sen:1996ph}
S.~Sen, \emph{Average entropy of a subsystem},
  \href{https://doi.org/10.1103/PhysRevLett.77.1}{\emph{Physical Review
  Letters} {\bfseries 77} (1996) 1}
  [\href{https://arxiv.org/abs/hep-th/9601132}{{\ttfamily hep-th/9601132}}].

\bibitem{vivo_pato_16}
P.~Vivo, M.P.~Pato and G.~Oshanin, \emph{Random pure states: Quantifying
  bipartite entanglement beyond the linear statistics},
  \href{https://doi.org/10.1103/PhysRevE.93.052106}{\emph{Physical Review E}
  {\bfseries 93} (2016) 052106}
  [\href{https://arxiv.org/abs/1602.01230}{{\ttfamily 1602.01230}}].

\bibitem{wei2017proof}
L.~Wei, \emph{Proof of vivo-pato-oshanin's conjecture on the fluctuation of von
  neumann entropy},
  \href{https://doi.org/10.1103/physreve.96.022106}{\emph{Physical Review E}
  {\bfseries 96} (2017) 022106}
  [\href{https://arxiv.org/abs/1706.08199}{{\ttfamily 1706.08199}}].

\bibitem{hackl2020bosonic}
L.~Hackl and E.~Bianchi, \emph{Bosonic and fermionic gaussian states from
  {K}ähler structures}, {\emph{arXiv preprint arXiv:2010.15518} (2020) }
  [\href{https://arxiv.org/abs/2010.15518}{{\ttfamily 2010.15518}}].

\bibitem{windt2020local}
B.~Windt, A.~Jahn, J.~Eisert and L.~Hackl, \emph{Local optimization on pure
  gaussian state manifolds},
  \href{https://doi.org/10.21468/scipostphys.10.3.066}{\emph{arXiv preprint
  arXiv:2009.11884} (2020) }
  [\href{https://arxiv.org/abs/2009.11884}{{\ttfamily 2009.11884}}].

\bibitem{Peschel2003}
I.~Peschel, \emph{Calculation of reduced density matrices from correlation
  functions}, \href{https://doi.org/10.1088/0305-4470/36/14/101}{\emph{J. Phys.
  A} {\bfseries 36} (2003) L205}.

\bibitem{Peschel2009}
I.~Peschel and V.~Eisler, \emph{Reduced density matrices and entanglement
  entropy in free lattice models},
  \href{https://doi.org/10.1088/1751-8113/42/50/504003}{\emph{J. Phys. A}
  {\bfseries 42} (2009) 504003}.

\bibitem{kieburg2019multiplicative}
M.~Kieburg, P.J.~Forrester and J.R.~Ipsen, \emph{Multiplicative convolution of
  real asymmetric and real anti-symmetric matrices},
  \href{https://doi.org/10.1515/apam-2018-0037}{\emph{Advances in Pure and
  Applied Mathematics} {\bfseries 10} (2019) 467}
  [\href{https://arxiv.org/abs/1712.04916}{{\ttfamily 1712.04916}}].

\bibitem{Forrester_2010}
P.J.~Forrester, \emph{Log-Gases and Random Matrices ({LMS}-34)}, Princeton
  University Press (dec, 2010),
  \href{https://doi.org/10.1515/9781400835416}{10.1515/9781400835416}.

\bibitem{eisert2010colloquium}
J.~Eisert, M.~Cramer and M.B.~Plenio, \emph{Colloquium: Area laws for the
  entanglement entropy},
  \href{https://doi.org/10.1103/revmodphys.82.277}{\emph{Reviews of Modern
  Physics} {\bfseries 82} (2010) 277}
  [\href{https://arxiv.org/abs/0808.3773}{{\ttfamily 0808.3773}}].

\bibitem{nahum2017quantum}
A.~Nahum, J.~Ruhman, S.~Vijay and J.~Haah, \emph{Quantum entanglement growth
  under random unitary dynamics},
  \href{https://doi.org/10.1103/physrevx.7.031016}{\emph{Physical Review X}
  {\bfseries 7} (2017) 031016}
  [\href{https://arxiv.org/abs/1608.06950}{{\ttfamily 1608.06950}}].

\bibitem{bauer2017stochastic}
M.~Bauer, D.~Bernard and T.~Jin, \emph{Stochastic dissipative quantum spin
  chains (i): Quantum fluctuating discrete hydrodynamics},
  \href{https://doi.org/10.21468/scipostphys.3.5.033}{\emph{SciPost Phys}
  {\bfseries 3} (2017) 033} [\href{https://arxiv.org/abs/1706.03984}{{\ttfamily
  1706.03984}}].

\end{thebibliography}\endgroup

\end{document}